\documentclass{PoS}

\usepackage{amsmath}
\usepackage{amssymb}
\usepackage{fontenc}
\usepackage{times}
\usepackage{mathptmx}
\usepackage{graphicx}

\title{Vertex Sensitivity in the Schwinger-Dyson Equations of QCD}

\ShortTitle{Vertex Sensitivity in the SDEs of QCD}

\author{\speaker{D.J. Wilson}\\
         Argonne National Laboratory\\
         Argonne, Illinois 60439, USA.
         E-mail: \email{djwilson@anl.gov}}

\author{M.R. Pennington\\
        Thomas Jefferson National Accelerator Facility\\
        Newport News, Virginia 23606, USA.
        E-mail: \email{michaelp@jlab.org}}

\abstract{The non-perturbative gluon and ghost propagators in Landau gauge QCD are obtained using the Schwinger-Dyson equation approach. The propagator equations are solved in Euclidean space using Landau gauge with a range of vertex inputs. Initially we solve for the ghost alone, using a model gluon input, which leads us to favour a finite ghost dressing in the non-perturbative region. In order to then solve the gluon and ghost equations simultaneously, we find that non-trivial vertices are required, particularly for the gluon propagator in the small momentum limit. We focus on the properties of a number vertices and how these differences influence the final solutions. The self-consistent solutions we obtain are all qualitatively similar and contain a masslike term in the gluon propagator dressing in agreement with related studies, supporting the long-held proposal of Cornwall.}

\FullConference{International Workshop on QCD Green's Functions, Confinement and Phenomenology,\\
		September 05-09, 2011\\
		Trento Italy}

\newcommand{\Gl}{\ensuremath{\mathcal{G}\ell}}
\newcommand{\Gh}{\ensuremath{\mathcal{G}h}}

\begin{document}        

\section{Introduction}

The non-perturbative gluon and ghost propagators may be obtained from their Schwinger-Dyson Equations (SDEs). They are interesting quantities that provide important information when studying confinement and dynamical chiral symmetry breaking in Quantum Chromodynamics (QCD). They are necessary inputs for studying bound states and amongst other things, can be used to obtain non-perturbative predictions for the running coupling $\alpha_s$. A number of early studies~\cite{Brown:1988bm,Brown:1988bn,Atkinson:1997tu,Bloch:2003yu,Alkofer:2003jr} found various singularities in the vanishing $p^2$ limit for the non-perturbative propagator dressings. However, more recent results from a variety of methods have found the dressings to be finite~\cite{Cucchieri:2007md,Aguilar:2008xm,Fischer:2008uz,Bogolubsky:2009dc,Binosi:2009qm,Cucchieri:2011ga,Cucchieri:2011um}.

In a recent study \cite{Pennington:2011xs}, the gluon and ghost propagator dressings, in the absence of quarks, were studied. The SDEs were solved self-consistently using a range of approximations for the vertices. Emphasis was given to how well these quantities are determined -- particularly in the physical region around 0.1 and 1 GeV. It turns out that the vertices are indeed very important if numerically precise results are desired.      

Landau gauge is used, since for this kind of problem it is the most appealing theoretically. The non-perturbative corrections to the ghost-gluon vertex are simplest in Landau gauge and statements regarding confinement are typically formulated in Landau gauge also. Furthermore, there are several lattice results to which we may compare.

The dressing functions that we solve for are related directly to the propagators, which for the gluon reads,
\begin{equation}
D_{\mu\nu}(p)= \frac{\Gl(p^2)}{p^2}\Bigl(g_{\mu\nu}-\frac{p_\mu p_\nu}{p^2}\Bigr)
\label{eq_gluonprop}
\end{equation}
and similarly for the ghost,
\begin{equation}
D(p)=-\frac{\Gh(p^2)}{p^2}
\end{equation}
where $D_{\mu\nu}(p)$ is the full gluon propagator in Landau gauge and $D(p)$ is the ghost propagator. The dressing functions $\Gl$ and $\Gh$ contain all of the non-perturbative physics of these two Green's functions. The simple transverse dressing of the gluon in Eq.~(\ref{eq_gluonprop}) is due to the Slavnov-Taylor identity for the propagator and is an important feature of the gauge invariance of the theory.

First we present the solutions of the ghost equation that are obtainable using a fixed gluon input. These solutions then provide a natural starting point for investigating the coupled system where simultaneous solutions of both propagator dressings are found. During the course of this work it became apparent that many simple vertices do not admit self-consistent solutions. We test a range of vertices with a range of solutions. A preferred system motivated by theoretical arguments is then selected. We conclude by comparing to the extant lattice data.
 
\section{The Ghost Equation}

Using a fixed gluon input we may solve the Ghost SDE alone and investigate its sensitivity to a range of input vertices. This is useful because this type of SDE is very simple to solve, and it teaches us what to expect when solving the more complicated, coupled equations self-consistently. In particular, the choice of ghost-gluon vertex applied here does have an effect, and understanding this is useful when solving the coupled system of gluons and ghosts.

In order to study the ghost equation in isolation we are required to provide a gluon input, we use the following model~\cite{Aguilar:2010gm},
\begin{equation}
\Gl(p^{2})=\frac{p^{2}}{m^{2}+p^{2}\left(1+\frac{11}{12\pi}\frac{N_{c}g^{2}}{4\pi}\mathrm{Log}\left(\frac{p^{2}+ m^2}{\mu^2}\right)\right)^{\frac{13}{22}}}\,.
\label{eq_gluonmodel}
\end{equation}
In this form the gluon contains a mass term $m$, which is typically $\mathcal{O}(\Lambda_{\mathrm{QCD}})$ in lattice and other SDE studies. The remaining terms arise in order to reproduce the one-loop behaviour. This is similar to many contemporary representations of the gluon dressing function, and a closely related form has recently been applied to a phenomenological study of hadron physics observables~\cite{Qin:2011dd,Qin:2011xq}. This form reproduces the leading resummed logarithm one finds from perturbation theory at large momenta.

\begin{figure}[tbh]
  \begin{center}
  \includegraphics[width=0.2\textwidth]{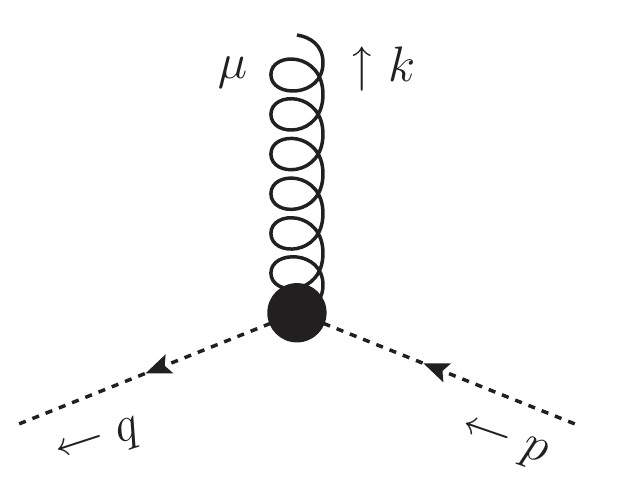}   
  \end{center}
  \caption{The ghost-gluon vertex indicating the momentum definition we adopt, the outgoing ghost momentum is $q$ and the gluon momentum is $k$.}
  \label{fig_gccdiag}
\end{figure}

The next step is to specify the ghost-gluon vertex which in principle is all that is required in order to solve the ghost equation given a fixed gluon input. The full ghost-gluon vertex has the form,
\begin{equation}
\Gamma_\mu=igf^{abc}\left(\alpha(k,p,q) q_\mu + \beta(k,p,q) k_\mu \right) 
\end{equation}
where $\alpha(k,p,q)$ and $\beta(k,p,q)$ are the non-perturbative dressing functions and the momenta are defined in Fig.~\ref{fig_gccdiag}. Initially we make use of the bare vertex in place of the fully dressed one, where $\alpha=1$ and $\beta=0$. This is motivated by Taylor's theorem~\cite{Taylor:1971ff} which constrains the sum of the functions to unity at vanishing incoming ghost momentum: $p=0$. When solving for the gluon, the following extended vertices will be used,
\begin{equation}
\Gamma_{\mu}^{(1)}=igf^{abc}\left(q_{\mu} - k_{\mu}\frac{k.q}{k^{2}}\mathcal{F}_\mathrm{IR}(k,p,q)\right),
\label{eq_gcctransv}
\end{equation}
\begin{equation}
\Gamma_{\mu}^{(2)}=igf^{abc}\left(q_{\mu} - p_{\mu}\frac{k.q}{k^{2}}\mathcal{N}_{\mathrm{IR}}\mathcal{F}_{\mathrm{IR}}(k,p,q)\right),
\label{eq_gcctt}
\end{equation}
where $\mathcal{F}_{\mathrm{IR}}=1$ for small momenta and $\mathcal{F}_{\mathrm{IR}}=0$ for large momenta. $\mathcal{N}_{\mathrm{IR}}$ is a non-perturbative normalisation parameter. 

The ghost-gluon vertex in Eq.~(\ref{eq_gcctransv}) is defined to be transverse in the small-$k$ limit, and has been used to solve these equations previously~\cite{Fischer:2008uz}. In the ghost equation, only the $q_\mu$ term survives and hence this gives identical results to the bare vertex when solving for the ghost alone. This is not the case for $\Gamma_{\mu}^{(2)}$, which is a minor modification that has been chosen in order to satisfy the Taylor condition. The additional term now vanishes when $p\to 0$, but contributes in both equations.

The final requirement is then to specify the renormalisation procedure. In the numerical method adopted here, it is most straightforward to use the momentum subtraction scheme, where the equations are subtracted from themselves at an arbitrary value of the propagator momentum. There are two choices that influence which solutions may be obtained. A subtraction at $p^2=0$ may produce a singular ghost dressing function. However a subtraction at some perturbative momentum can be applied and in some ways is better since the dressings are already well known in that region. The Schwinger-Dyson equations tell us how these evolve from the known results at large momenta into the non-perturbative region. The alternative is to make an assumption about the ghost dressing at zero momentum.

\begin{figure}[tbh]
  \begin{center}
  \includegraphics[width=0.45\textwidth]{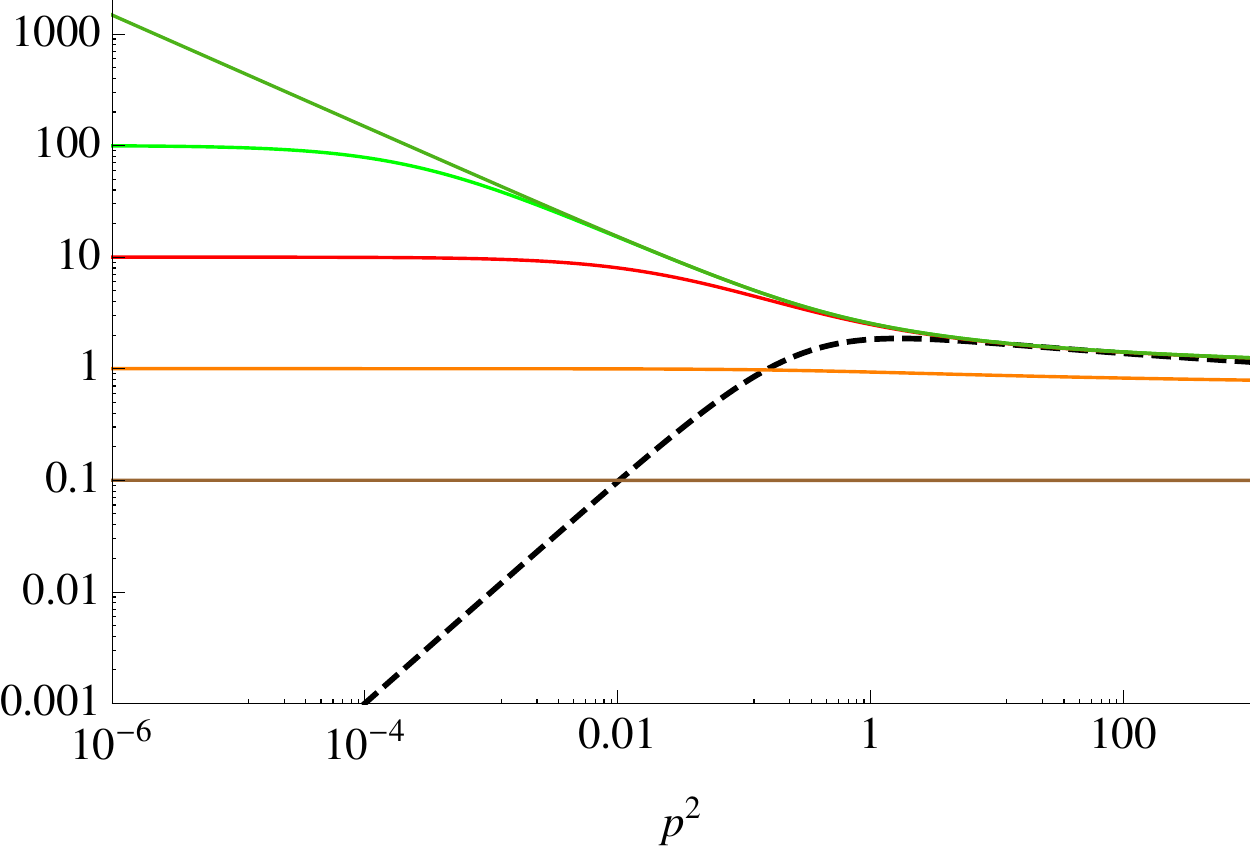}   
  \end{center}
  \caption{Examples of ghost solutions on a log-log plot subtracting at zero momentum.  The solid curves display different ghost dressings $\Gh(p^{2})$,
while the dashed curve is the gluon dressing $\Gl(p^{2})$. Only the specified subtraction value $\Gh(0)$ is varied between the solutions which can be read from the plot. The $\Gh(p^{2}=0)$ value is fixed and the ghost equation solved with the depicted gluon until the ghost inputs and outputs are self-consistent.  The units of $p^{2}$ are arbitrary since we have not fixed the coupling to the physical value, but may be considered to be ${\cal O}\,(1$ GeV$^2$).}
  \label{fig_mixedghostsols}
\end{figure}

The renormalised ghost equation with a cutoff regularisation reads,
\begin{align}
\Gh^{-1}(p^2)&=\tilde{Z_3}(\mu^2,\Lambda^2) + \Pi_{gc}(p^2,\mu^2)\\
             &=\Gh(\mu^2) + \Pi_{gc}(p^2,\mu^2) + \Pi_{gc}(\mu^2,\mu^2)
\end{align}
in the second line we eliminate the renormalisation constant $\tilde{Z}_3(\mu^2,\Lambda^2)$. The function $\Pi_{gc}(p^2,\mu^2)$ is the ghost self-energy diagram, the details of which are given in~\cite{Pennington:2011xs}. This is the equation that is solved numerically.
To check the dependence and range of solutions available, we vary the value of $1/\Gh(0)$. The solutions are given in Fig.~\ref{fig_mixedghostsols}. We observe two classes of solutions. With $\Gh(0)\gtrsim2$ then the Ghost dressings in the perturbative region are practically indistinguishable. In the small $p^2$ region any value in the range $\infty>\Gh(0)>2$ produces the same perturbative solution. Reducing $\Gh(0)$ there is then some critical value where the solutions in the perturbative region also begin to fall. The precise value of $\Gh(0)$ for which this occurs is dependent upon the couplings, the vertex and the gluon. All of the solutions are monotonic and decrease with increasing $p^2$. By specifying the value at $p^2=0$ then all other points are implicitly given by the equation.

We may then switch to the other viewpoint of subtracting in the perturbative region and choose the typical perturbative condition $\Gh(\mu^2)=1$. In doing this, none of the solutions subtracted at zero in Fig.~\ref{fig_mixedghostsols} are obtained. The ghost dressing rises from unity at $p^2=\mu^2$ to a finite value at zero momentum. It is possible to subtract at zero and fine tune to this value and obtain the same solution, although this is a special case.

\begin{figure}[tbh]
  \begin{center}
  \includegraphics[width=0.45\textwidth]{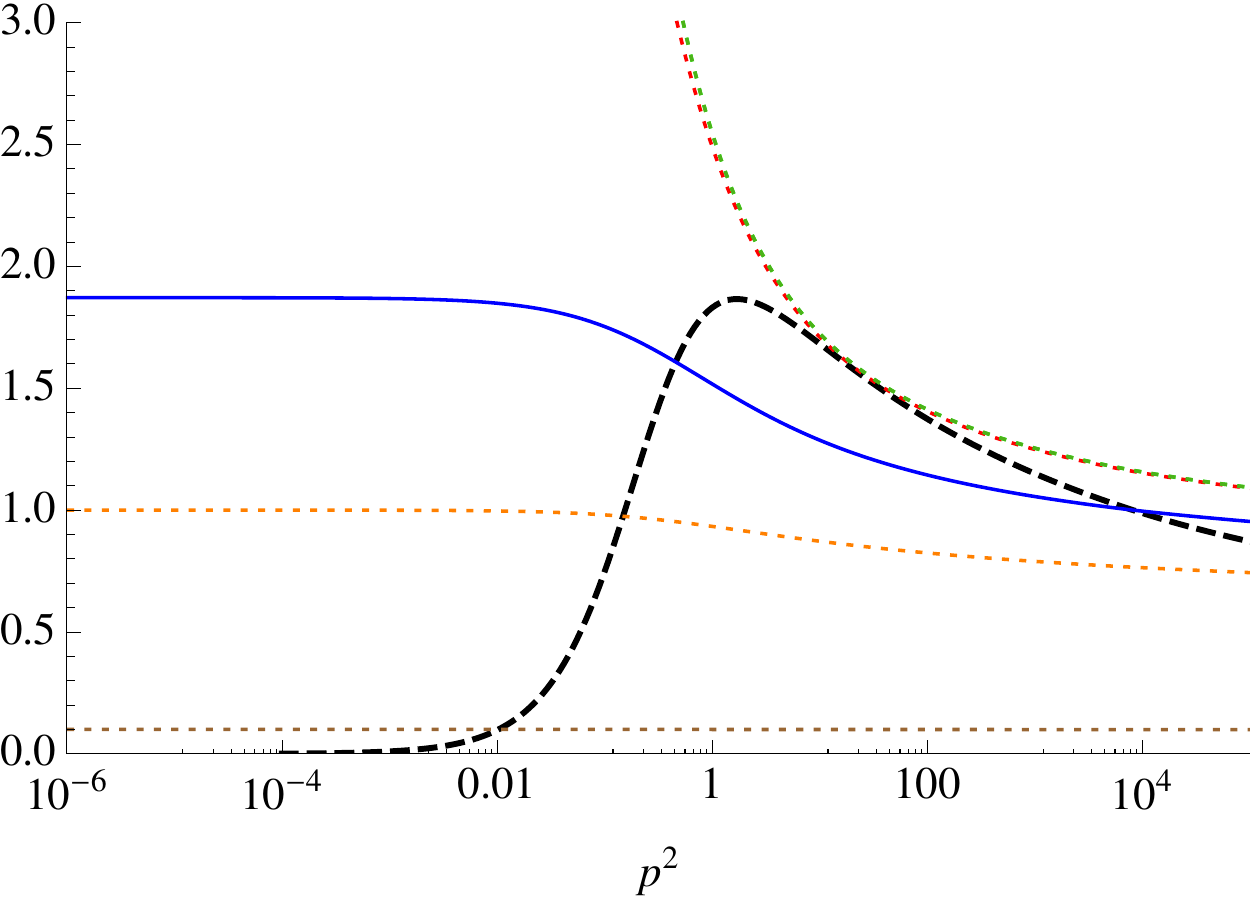}   
  \end{center}
  \caption{The dotted curves correspond to the zero momentum subtracted solutions and the colours match Fig.~5. The dashed curve is the gluon input, $\Gl(p^2)$, and the solid curve is the physically relevant solution of the ghost equation, $\Gh(p^2)$. The units of $p^2$ are arbitrary.}
  \label{fig_ghostlin}
\end{figure}

Specifying $\Gh(\mu^2)=1$ as one would do perturbatively leads to the interesting consequence that only one of these many Ghost dressings is obtained. This is distinct from singular solution and importantly no renormalisation group running connects this solution to the singular one. Since the expectation is that asymptotically free QCD at large momentum transfers is accurately described by the perturbative solution, this is the solution we favour. This effect is depicted in Fig.~\ref{fig_ghostlin}, where the solid curve is the unique solution found by imposing $\Gh(\mu^2)=1$ whilst the remaining curves are those subtracted at zero from Fig.~\ref{fig_mixedghostsols}.

Hence we carry forward this arrangement when solving the gluon equation. It is worth noting that this same effect has been tested for gluons that vanish more or less quickly and the same qualitative features arise. Similar effects have been observed in other studies~\cite{Boucaud:2008ky,Pene:2009iq,RodriguezQuintero:2010wy,Watson:2010cn}.

 \begin{figure}[tbh]
   \begin{center}
   \includegraphics[width=0.45\textwidth]{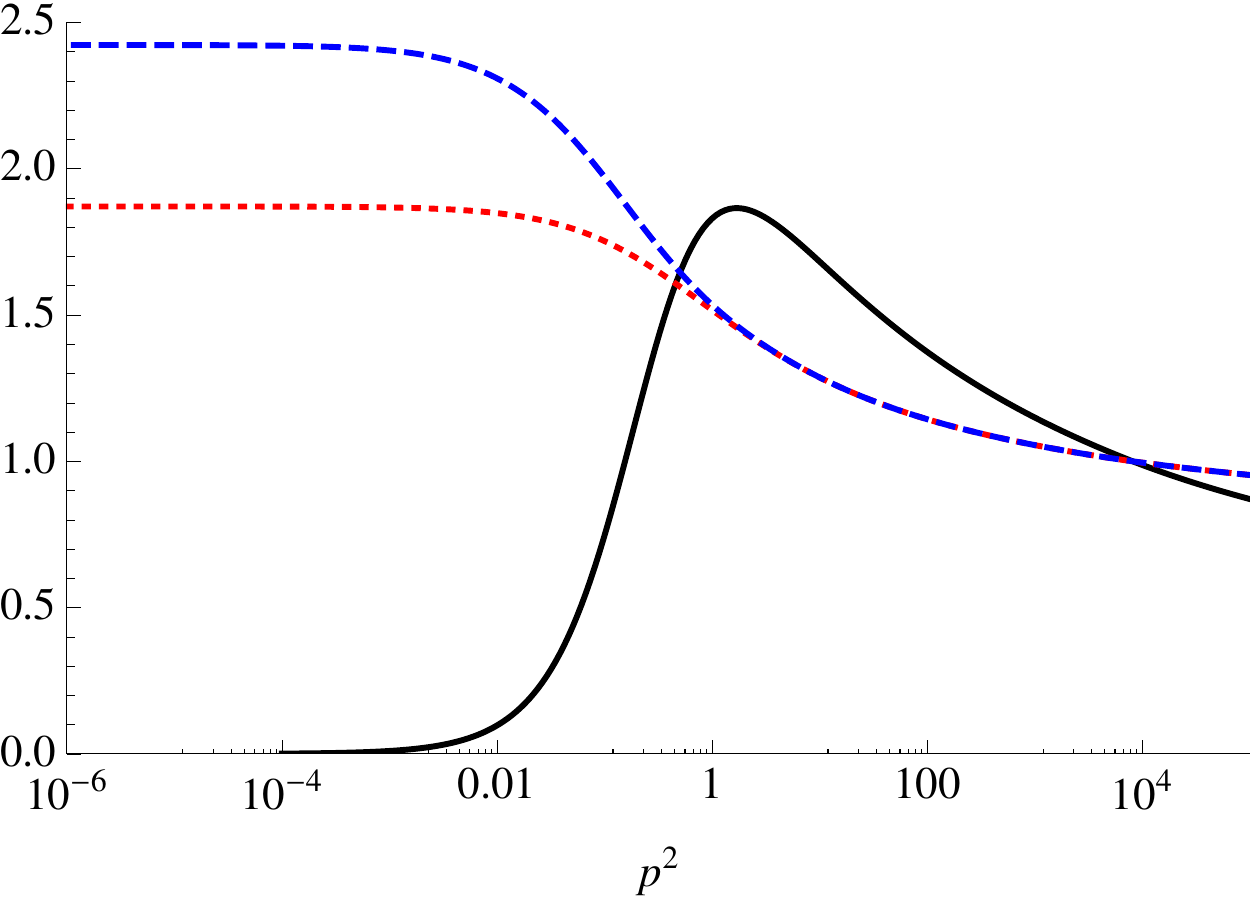}   
   \end{center}
   \caption{A comparison of two solutions normalised using the perturbative condition. The (red) dotted curve is the ghost solution, $Gh(p^2)$, with a bare vertex or transverse vertex. The (blue) dashed curve is the ghost dressing obtained using $\Gamma^{(2)}_\mu$. The (black) solid curve is the fixed gluon dressing input $\Gl(p^2)$ from the model. The units of $p^2$ are arbitrary internal units.}
   \label{fig_ghosttay}
 \end{figure}

Finally we compare the result using $\Gamma_{\mu}^{(1)}$, which is identical to the bare vertex in the ghost equation, to the solution found using $\Gamma_{\mu}^{(2)}$. This comparison is shown in Fig.~\ref{fig_ghosttay} where a large difference is apparent between the two results in the non-perturbative region. 

\section{The coupled Gluon and Ghost system}

The coupled system is naturally more complicated, although we already have most of the ingredients that are needed. We choose to work using only the one-loop graphs since the two-loop graphs are a considerable complication numerically. It is possible that these terms are important and in order to obtain numerically precise results it is likely that they are essential. The gluon equation has the form
\begin{align}
\Gl^{-1}(p^2)&=Z_3(\mu^2,\Lambda^2) + \Pi_{2c}(p^2,\mu^2) + \Pi_{2g}(p^2,\mu^2)\\
             &=\Gl^{-1}(\mu^2)+\Pi_{2c}(p^2,\mu^2)-\Pi_{2c}(\mu^2,\mu^2) + \Pi_{2g}(p^2,\mu^2) -\Pi_{2g}(\mu^2,\mu^2)\label{eq_gluonsub}
\end{align}
where in the second line we have subtracted to remove the renormalisation constant, $\Pi_{2g}$ is the gluon loop contribution to the vacuum polarisation, and $\Pi_{2c}$ arises due to the ghost loop. The ghost-gluon vertices we have already specified are sufficient for $\Pi_{2c}$. The final requirement is to specify the triple-gluon vertex. To illustrate the uncertainty we select three different vertices and check the sensitivity in the solutions.

The first we apply is a simple form that reproduces the resummed leading logarithmic running found at one-loop order in perturbation theory~\cite{Bloch:2003yu},
\begin{align}
\Gamma^{(A)}_{\mu\nu\rho}(k,p,q) = \frac{\Gh(p^2)\Gh(q^2)}{\Gl(p^2)\Gl(q^2)}\Gamma^{(0)}_{\mu\nu\rho}(k,p,q)
\end{align}
this is proportional to the bare Lorentz structure $\Gamma^{(0)}_{\mu\nu\rho}(k,p,q)$ and $p$ and $q$ are chosen to be the internally contracted legs under the loop integration. What one immediately notices about this form is that it is not symmetric under interchange of legs. Since this vertex is expected to be symmetric to all orders and since the Ward-Slavnov-Taylor Identity (WSTI) is also symmetric under the interchange of legs, a symmetric form is desirable. Hence we also apply,
\begin{align}
\Gamma^{(B)}_{\mu\nu\rho}(k,p,q)&=\Gamma^{(0)}_{\mu\nu\rho}(k,p,q)\,
\frac{1}{3}\left(\frac{\Gh(k^{2})}{\Gl(k^{2})}+\frac{\Gh(p^{2})}{\Gl(p^{2})}+\frac{\Gh(q^{2})}{\Gl(q^{2})}\right)
\end{align}
where the ghost-over-gluon form is suggested by the WSTI, and for simplicity this is chosen to be proportional to the bare Lorentz structure. The final dressing of the triple-gluon vertex we apply is a solution of the WSTI using a bare ghost-gluon scattering kernel,
\begin{align}
\Gamma^{(C)}_{\mu\nu\rho}(k,p,q)&=\frac{1}{2}\left(\frac{\Gh(q^{2})}{\Gl(p^{2})}+\frac{\Gh(q^{2})}{\Gl(k^{2})}\right)g_{\mu\nu}(k-p)_{\rho}\nonumber\\
                                &+\frac{1}{2}\left(\frac{\Gh(k^{2})}{\Gl(q^{2})}+\frac{\Gh(k^{2})}{\Gl(p^{2})}\right)g_{\nu\rho}(p-q)_{\mu}\nonumber\\
                                &+\frac{1}{2}\left(\frac{\Gh(p^{2})}{\Gl(k^{2})}+\frac{\Gh(p^{2})}{\Gl(q^{2})}\right)g_{\rho\mu}(q-k)_{\nu}\; ,
\label{eq_gggwsti}
\end{align}
now each term of the bare structure receives its own non-perturbative dressing and these have a symmetric structure. More elaborate forms are also possible and eventually something more will be required in order to obtain a triple-gluon vertex that has a consistent ghost-gluon scattering kernel with the ghost-gluon vertex as prescribed by the WSTI.

\begin{figure}[tbh]
  \begin{center}
  \includegraphics[width=0.45\textwidth]{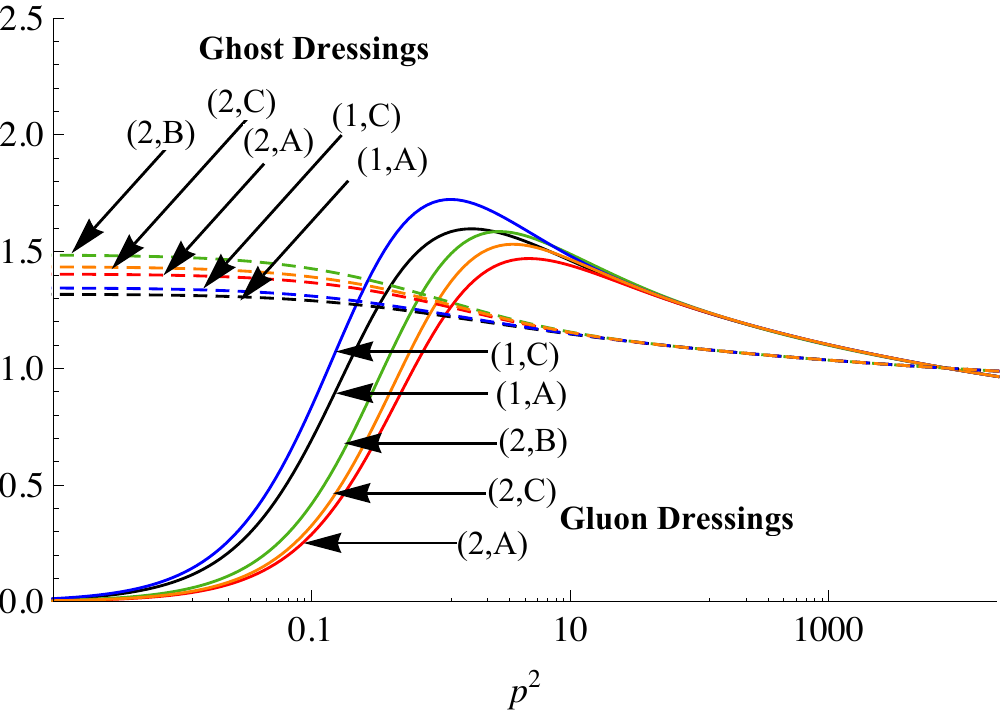}   
  \end{center}
  \caption{The range of solutions obtained using the possible vertex combinations. The label $(i,j)$ refers to the vertices used in obtaining the solutions corresponding to $\Gamma^{(i)}_{\mu}$ for the ghost-gluon vertex and $\Gamma^{(j)}_{\mu\nu\rho}$ for the triple-gluon vertex. The missing curve corresponds to the $\Gamma^{(1)}_{\mu}$ and $\Gamma^{(B)}_{\mu\nu\rho}$, self-consistent solutions were not obtained there. The parameters are not varied between the solutions, only the vertices. The units of $p^{2}$ are arbitrary since we have not fixed the coupling to a physical value.}
  \label{fig_vertices}
\end{figure}

The next step is to implement these vertices numerically in the ghost and gluon SDEs and to solve them self-consistently. This is done numerically in the momentum subtraction scheme, subtracting both equations at the renormalisation point $\mu$. The equations are solved iteratively and are fairly insensitive to the initial function. The model gluon in Eq.~(\ref{eq_gluonmodel}) may be used and $\Gh(p^2)=1$ is adequate for the ghost. Convergence is achieved rapidly using a Newton-Raphson method, although a natural iterative procedure is sufficient. 

In Fig.~\ref{fig_vertices} we see a broad range of gluon dressings given the different vertex inputs. The position of the peak of the gluon propagator moves around considerably. This is important because it is related to the effective gluon mass and physical quantities depend on this, as has been shown in~\cite{Qin:2011dd,Qin:2011xq}. This plot tells us that just using any old vertex is insufficient and one should appeal to all possible constraints when choosing this input.

These differences in the propagator dressings arise from differences in the vacuum polarisation integrals, which come directly from the vertices. It is informative to look inside the contributions for the gluon from each loop to disentangle the range of vertex contributions. The simplest triple-gluon vertex $\Gamma^{(A)}_{\mu\nu\rho}$ is actually quite close to the completely bare triple-gluon vertex in the small $p^2$ region, whilst the dressed symmetric triple-gluon vertices give quite different results. 

\begin{figure}[!hbt]
  \begin{center}
  \includegraphics[width=0.45\textwidth]{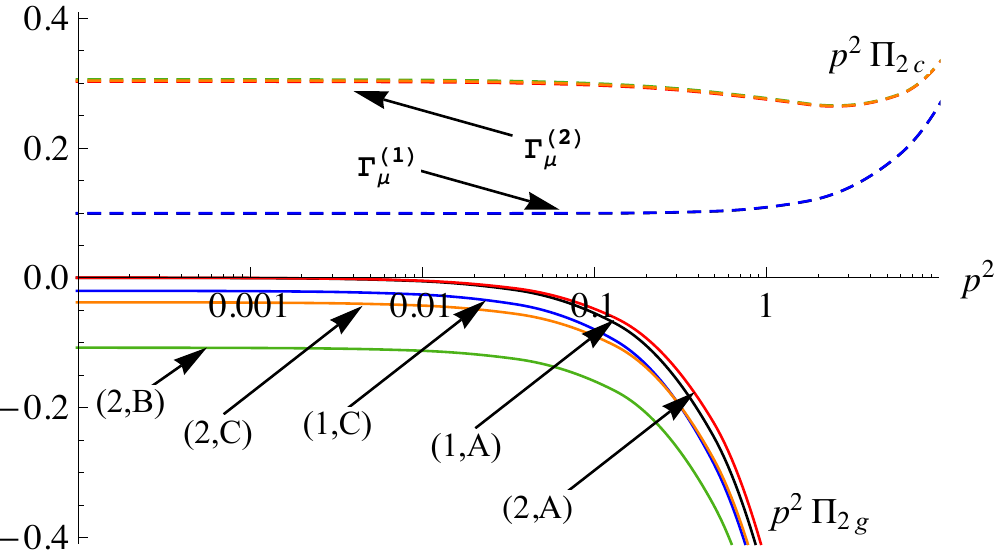}  
  \includegraphics[width=0.45\textwidth]{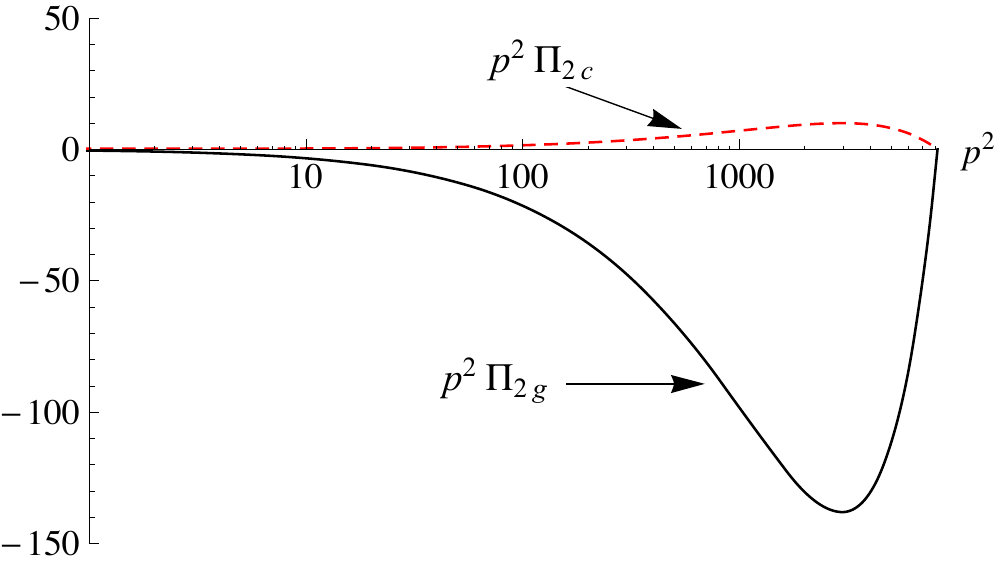}    
  \end{center}
  \caption{Gluon polarization functions, $p^{2}\Pi_{2c}(p^{2})$ (positive values) and $p^{2}\Pi_{2g}(p^{2})$ (negative values). Left: IR region, Right: UV region. In the UV the curves are indistinguishable on this scale. In the IR, the ghost-loop curves for multiple solutions lie on top of each other. The bare ghost-gluon vertex, not shown here, always gives a vanishing IR contribution given an IR vanishing gluon dressing and an IR finite ghost dressing. The functions $p^{2}\Pi_{2g}(p^{2},\mu^{2})$ are labelled according to the input vertices with the contributions from $\Gamma_{\mu}^{(i)}$ and $\Gamma_{\mu\nu\rho}^{(j)}$ labelled as $(i,j)$ in the plot. The units of $p^{2}$ are arbitrary.
  }
  \label{fig_pols}
\end{figure}

\noindent In order to see this we take the subtracted contribution from each loop integration and since they typically behave as $1/p^2$ for small $p^2$, we multiply them by $p^2$ in order to get a better view of the differences. This we show in Fig.~\ref{fig_pols}, where the quantities plotted are $p^2\left(\Pi_{i}(p^2,\mu^2)-\Pi_{i}(\mu^2,\mu^2)\right)$, from Eq.~\ref{eq_gluonsub}. In the right panel we see the large $p^2$ region, which is essentially determined by perturbation theory, where all contributions are the same. Differences arise in the left panel of Fig.~\ref{fig_pols} where we zoom in on the small $p^2$ region.

The negative curves are the contributions from the triple-gluon vertices. The plotted functions $p^{2}\Pi_{2g}(p^{2},\mu^{2})$ are labelled according to the input vertices with the contributions from $\Gamma_{\mu}^{(i)}$ and $\Gamma_{\mu\nu\rho}^{(j)}$ labelled as $(i,j)$ in the plot. It is evident that $\Gamma^{(A)}_{\mu\nu\rho}$ leads to a vanishing contribution when plotted in this way whilst the sensibly dressed symmetric vertices do not, they are finite and negative. The gluon dressing function is not expected to change sign and is positive in the perturbative region, so it is necessary that a positive contribution also arises to cancel that from the gluon loop. In the one-loop-only system with these triple-gluon vertices, a positive contribution must arise from the ghost-loop terms. This is indeed the case since we have self-consistent solutions for the curves that are plotted. The combination $(1,B)$ is missing since for this set of parameters it did not satisfy the 
 condition. It is then not possible to find self-consistent solutions. The same condition is also why we need non-trivial ghost-gluon vertices, since the bare vertex gives a vanishing contribution when plotted in this way. This leads directly to a change of sign in the gluon dressing and no stable set of solutions can be found.


The fact that a symmetric triple-gluon dressing results in a negative contribution is very important, since it leads to the condition that the ghost loop, and hence the ghost-gluon vertex, need not be transverse alone and can work together with the other loops, as they do in the perturbative region, in order to produce a transverse gluon dressing.

\section{Comparison with Lattice QCD}

Primarily we have been motivated by theoretical issues encountered in solving the Schwinger-Dyson equations for the gluon and ghost propagators. However lattice studies exist for these quantities and the results are complementary. The possible issues that result from a applying a finite grid to Euclidean spacetime are quite different to those that we may induce here by truncation, so a comparison is a useful independent cross-check. Importantly there are lattice computations that are precise and in the pure gauge sector.

An early lattice result favoured a finite, massive solution of the gluon~\cite{Mandula:1987rh}. Following this many recent lattice studies now exist using very large lattices~\cite{Cucchieri:2007md,Cucchieri:2011ga,Cucchieri:2011um,Bicudo:2010wi}. We compare our calculations to \cite{Bogolubsky:2009dc}, which provides results in Landau gauge for both dressing functions. The qualitative behaviour is the same, with a finite ghost dressing function and a vanishing gluon propagator dressing function.    

\begin{figure}[thb]
  \begin{center}
  \includegraphics[width=0.45\textwidth]{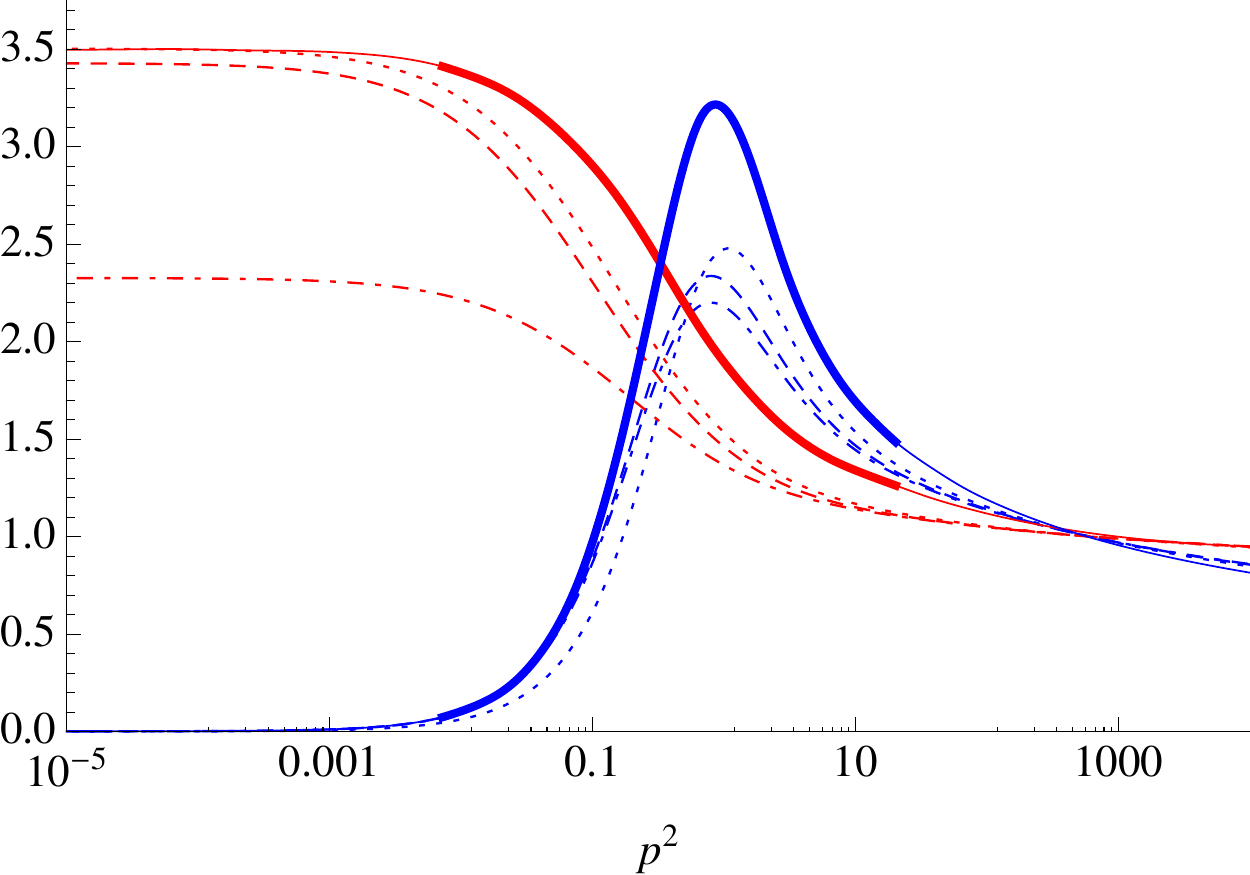}   
  \includegraphics[width=0.45\textwidth]{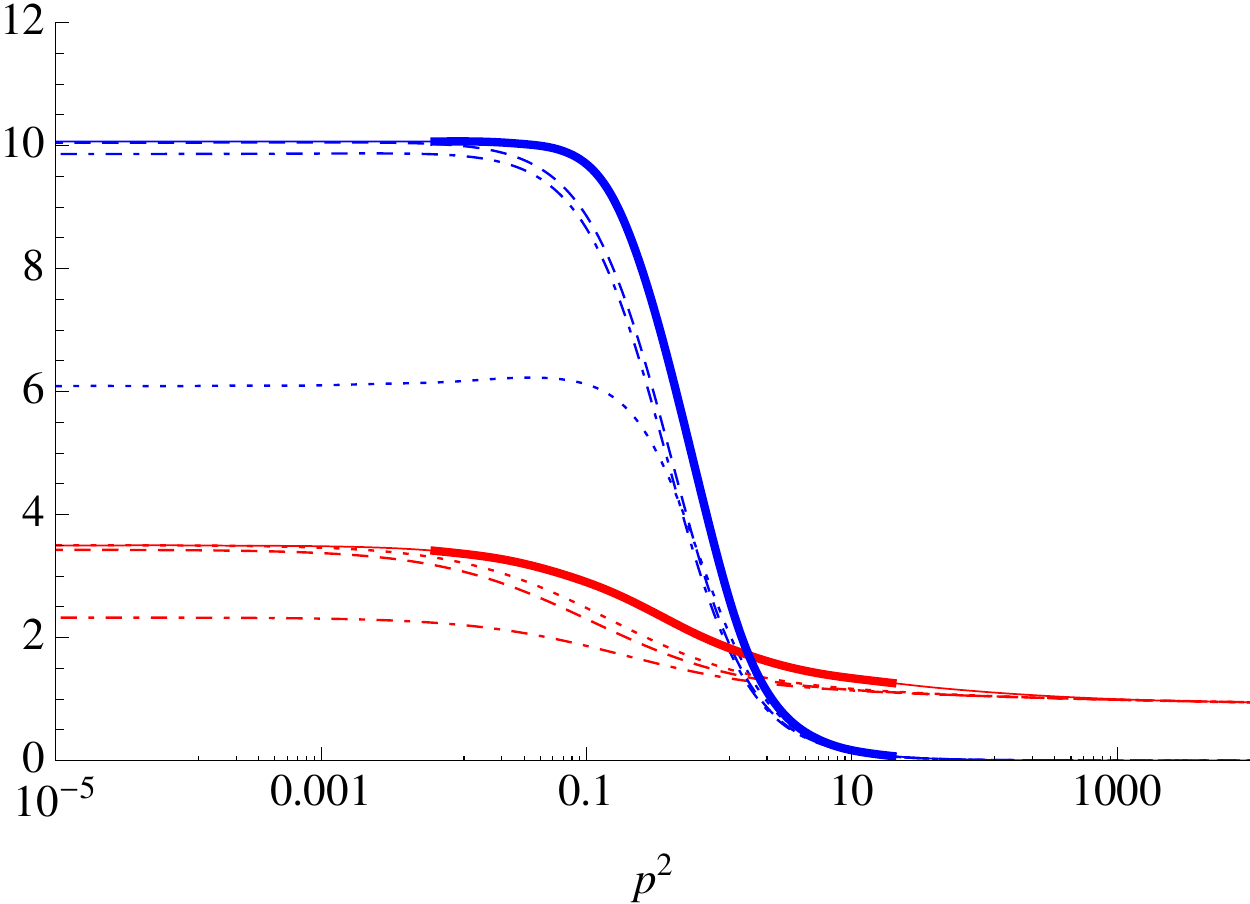}
  \end{center}
  \caption{Self-consistent solutions tuned to lattice solutions, showing the dressing functions. The solid curves depict a smooth fit to the lattice data. The heavier region is where the functions are represented by lattice data and the feint region represents the natural extrapolation. The broken curves are the tuned solutions. Left: The blue curve corresponds to the gluon dressing function $\Gl(p^{2})$, this vanishes as $p^{2}\to0$, the red curve corresponds to the ghost dressing function $\Gh(p^{2})$. Right: The blue curve corresponds to $\Gl(p^{2})/p^{2}$, this is $\sim10$ as $p^{2}\to0$, the red curve corresponds to the ghost dressing function again.}
  \label{fig_latdf}
\end{figure}

In Fig.~\ref{fig_latdf} we show the solutions of the equations obtained self-consistently with three different sets of parameters that reproduce different features of the lattice data. In obtaining these solutions we use the preferred set of vertices, $\Gamma^{(C)}_{\mu\nu\rho}$ from Eq.~(\ref{eq_gggwsti}) and the ghost-gluon vertex given in Eq.~(\ref{eq_gcctt}). We then tune the parameters to obtain a reasonable representation of the lattice data. We are not able to find a close fit over the whole momentum region. All three sets of parameters fail to reproduce the magnitude of the peak of the gluon propagator dressing function shown in the left panel of Fig.~\ref{fig_latdf}. It is possible that this is due to the effects of the two-loop graphs. However to be sure one would need to  perform the full dressed two-loop integrals.

Another difference is the value of the coupling in the perturbative region. In all three sets of solutions, the lattice appears to have a larger value of the coupling. This leads to a slower increase in the SDE solutions between the peak of the gluon propagator and the subtraction point. A similar effect is seen in the ghost dressings in this region. In the non-perturbative region it is possible to find similar values of the dressings simultaneously: The dashed solution is the closest we show and the effect is most visible in the right panel of Fig.~\ref{fig_latdf}.

\section{Conclusion}

We solved the Schwinger-Dyson Equations of QCD in the Landau gauge in the absence of quarks. In solving the ghost equation we found the singular ghost is excluded when a subtraction point in the perturbative region is selected. This was true for all vertices that we tested and for a range of gluon inputs although only one was shown. This agrees with earlier findings~\cite{Boucaud:2008ky}.

\begin{figure}[tbh]
 \begin{center}
  \includegraphics[width=0.45\textwidth]{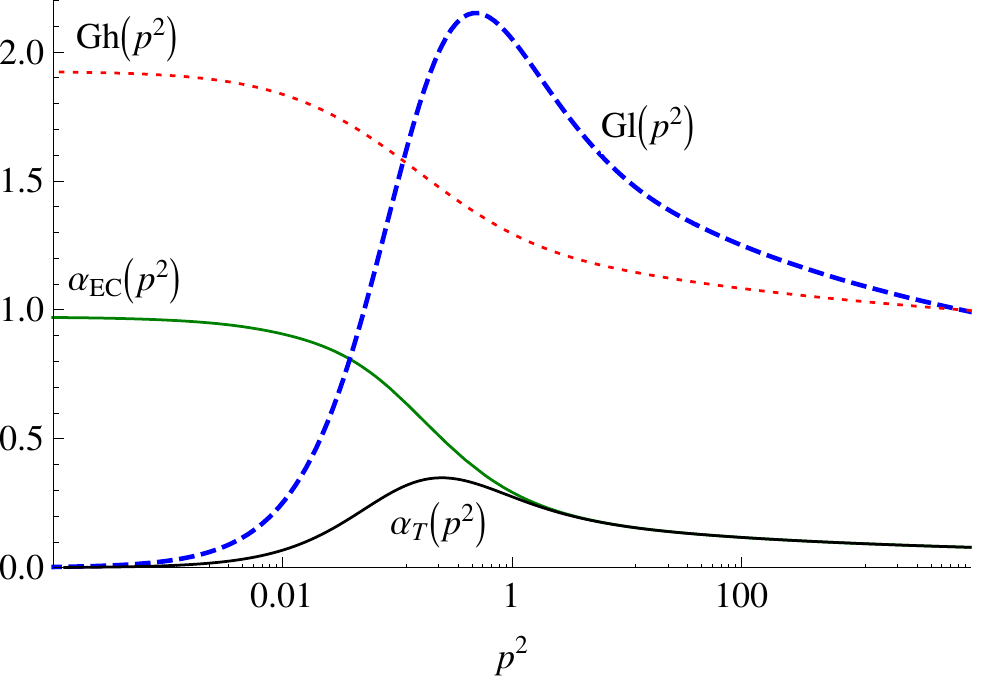}   
  \end{center}
  \caption{The running coupling $\alpha_{T}(p^{2})$ is shown as the black solid curve. The running coupling $\alpha_{EC}(p^{2})$ is shown as the green solid curve. The gluon dressing $\Gl(p^{2})$ is the dashed blue curve and the dotted red curve is the ghost dressing function $\Gh(p^{2})$. The units of $p^{2}$ are now very close to GeV$^{2}$.}
  \label{fig_rc}
\end{figure}

When solving the gluon equation we found that non-trivial vertex dressings were essential in that simple vertices did not lead to self-consistent solutions. It was found that symmetric triple-gluon vertex dressings lead generically to a non-zero contribution of the gluon-loop diagram in the non-perturbative region. The main implication of this is that the ghost-loop term is not required to be transverse alone and that both terms can contribute to the transversality of the gluon propagator, as they do in the perturbative region. This combined with Taylor's theorem led us to test a simple modification of $\Gamma^{(1)}_{\mu}$, a vertex already known to give self-consistent solutions with $\Gamma^{(2)}_{\mu}$ in~Eq.~(\ref{eq_gcctt}). This modification protects Taylor's theorem whilst sacrificing the transversality of that individual loop. It is interesting to note that another ghost-gluon vertex dressing presented at this workshop leads to similar effects in the ghost equation~\cite{Boucaud:2011eh}.

A range of solutions were found by selecting a conservative set of parameters and testing the effects of using the different vertices. A wide variation was found in the physical region leading us to the suggestion that more advanced vertices are required in order to obtain numerically precise results. Proposals for new vertices in a related method specifically investigating the mass generation mechanism have been suggested~\cite{Papavassiliou:2011ty,Aguilar:2011xe}.

We may use the method we present here to produce a running coupling by considering the ghost-gluon vertex renormalisation. There are two definitions we apply, first using the standard Taylor definition which we denote $\alpha_T$ and a second form, $\alpha_{EC}$ related to the Taylor form, but removes the mass term from the gluon dressing~\cite{Aguilar:2009nf}. These curves and a set of solutions fixed using physically meaningful parameters are given in Fig.~\ref{fig_rc}. The effective gluon mass we found from this gluon propagator was $m_g=400$ MeV.

These equations produced self-consistent solutions of the ghost and gluon propagator dressings that were in qualitative agreement with lattice QCD results. Quantitative differences from this method could arise in either the vertex dressings or two-loop graphs. Regardless of precision, qualitatively all results are the same and they point to a finite gluon propagator in the non-perturbative region, corresponding to a dynamically generated gluon mass as proposed by Cornwall~\cite{Cornwall:1981zr}.

\section*{Acknowledgements}
The Institute for Particle Physics Phenomenology at Durham University, UK, its staff and students are gratefully aknowledged for providing an ideal working environment for much of this study.  DJW thanks  Jefferson Laboratory for hospitality while finalising this work. This paper has in part been authored by Jefferson Science Associates, LLC under U.\ S.\ DOE Contract No. DE-AC05-06OR23177. 
This work was also supported by the U.\ S.\ Department of Energy, Office of Nuclear Physics, Contract No.~DE-AC02-06CH11357. 

\bibliographystyle{jhep}
\bibliography{sderefs2}

\providecommand{\href}[2]{#2}\begingroup\raggedright\begin{thebibliography}{10}

\bibitem{Brown:1988bm}
N.~Brown and M.~R. Pennington, {\it {Studies of confinement: how quarks and
  gluons propagate}},  {\em Phys. Rev.} {\bf D38} (1988) 2266.

\bibitem{Brown:1988bn}
N.~Brown and M.~R. Pennington, {\it {Studies of Confinement: How the Gluon
  Propagates}},  {\em Phys. Rev.} {\bf D39} (1989) 2723.

\bibitem{Atkinson:1997tu}
D.~Atkinson and J.~C.~R. Bloch, {\it {Running coupling in non-perturbative QCD.
  I: Bare vertices and y-max approximation}},  {\em Phys. Rev.} {\bf D58}
  (1998) 094036 [\href{http://arXiv.org/abs/hep-ph/9712459}{{\tt
  hep-ph/9712459}}].

\bibitem{Bloch:2003yu}
J.~C.~R. Bloch, {\it {Two-loop improved truncation of the ghost-gluon Dyson-
  Schwinger equations: Multiplicatively renormalizable propagators and
  nonperturbative running coupling}},  {\em Few Body Syst.} {\bf 33} (2003)
  111--152 [\href{http://arXiv.org/abs/hep-ph/0303125}{{\tt hep-ph/0303125}}].

\bibitem{Alkofer:2003jr}
R.~Alkofer, C.~S. Fischer, H.~Reinhardt and L.~von Smekal, {\it {On the
  infrared behaviour of gluons and ghosts in ghost- antighost symmetric
  gauges}},  {\em Phys. Rev.} {\bf D68} (2003) 045003
  [\href{http://arXiv.org/abs/hep-th/0304134}{{\tt hep-th/0304134}}].

\bibitem{Cucchieri:2007md}
A.~Cucchieri and T.~Mendes, {\it {What's up with IR gluon and ghost propagators
  in Landau gauge? A puzzling answer from huge lattices}},  {\em PoS} {\bf
  LAT2007} (2007) 297 [\href{http://arXiv.org/abs/0710.0412}{{\tt 0710.0412}}].

\bibitem{Aguilar:2008xm}
A.~C. Aguilar, D.~Binosi and J.~Papavassiliou, {\it {Gluon and ghost
  propagators in the Landau gauge: Deriving lattice results from
  Schwinger-Dyson equations}},  {\em Phys. Rev.} {\bf D78} (2008) 025010
  [\href{http://arXiv.org/abs/0802.1870}{{\tt 0802.1870}}].

\bibitem{Fischer:2008uz}
C.~S. Fischer, A.~Maas and J.~M. Pawlowski, {\it {On the infrared behavior of
  Landau gauge Yang-Mills theory}},  {\em Annals Phys.} {\bf 324} (2009)
  2408--2437 [\href{http://arXiv.org/abs/0810.1987}{{\tt 0810.1987}}].

\bibitem{Bogolubsky:2009dc}
I.~L. Bogolubsky, E.~M. Ilgenfritz, M.~Muller-Preussker and A.~Sternbeck, {\it
  {Lattice gluodynamics computation of Landau gauge Green's functions in the
  deep infrared}},  {\em Phys. Lett.} {\bf B676} (2009) 69--73
  [\href{http://arXiv.org/abs/0901.0736}{{\tt 0901.0736}}].

\bibitem{Binosi:2009qm}
D.~Binosi and J.~Papavassiliou, {\it {Pinch Technique: Theory and
  Applications}},  {\em Phys. Rept.} {\bf 479} (2009) 1--152
  [\href{http://arXiv.org/abs/0909.2536}{{\tt 0909.2536}}].

\bibitem{Cucchieri:2011ga}
A.~Cucchieri and T.~Mendes, {\it {Further Investigation of Massive Landau-Gauge
  Propagators in the Infrared Limit}},  {\em PoS} {\bf LATTICE2010} (2010) 280
  [\href{http://arXiv.org/abs/1101.4537}{{\tt 1101.4537}}].

\bibitem{Cucchieri:2011um}
A.~Cucchieri and T.~Mendes, {\it {The Saga of Landau-Gauge Propagators:
  Gathering New Ammo}},  {\em {in {\it The IXth International Conference on
  Quark Confinement and the Hadron Spectrum QCHS IX}, edited by F.J.
  Llanes-Estrada and J.R. Pelaez}, AIP Conf. Proc.} {\bf 1343} (2011) 185--187
  [\href{http://arXiv.org/abs/1101.4779}{{\tt 1101.4779}}].

\bibitem{Pennington:2011xs}
M.~R. Pennington and D.~J. Wilson, {\it {Are the Dressed Gluon and Ghost
  Propagators in the Landau Gauge presently determined in the confinement
  regime of QCD?}},  {\em Phys.Rev.} {\bf D84} (2011) 094028
  [\href{http://arXiv.org/abs/1109.2117}{{\tt 1109.2117}}].

\bibitem{Aguilar:2010gm}
A.~C. Aguilar, D.~Binosi and J.~Papavassiliou, {\it {QCD effective charges from
  lattice data}},  {\em JHEP} {\bf 1007} (2010) 002
  [\href{http://arXiv.org/abs/1004.1105}{{\tt 1004.1105}}].

\bibitem{Qin:2011dd}
S.-x. Qin, L.~Chang, Y.-x. Liu, C.~D. Roberts and D.~J. Wilson, {\it
  {Interaction model for the gap equation}},  {\em Phys.Rev.} {\bf C84} (2011)
  042202 [\href{http://arXiv.org/abs/1108.0603}{{\tt 1108.0603}}].

\bibitem{Qin:2011xq}
S.-x. Qin, L.~Chang, Y.-x. Liu, C.~D. Roberts and D.~J. Wilson, {\it
  {Commentary on rainbow-ladder truncation for excited states and exotics}},
  \href{http://arXiv.org/abs/1109.3459}{{\tt 1109.3459}}.

\bibitem{Taylor:1971ff}
J.~C. Taylor, {\it {Ward Identities and Charge Renormalization of the Yang-
  Mills Field}},  {\em Nucl. Phys.} {\bf B33} (1971) 436--444.

\bibitem{Boucaud:2008ky}
P.~Boucaud, J.~P. Leroy, A.~Le~Yaouanc, J.~Micheli, O.~Pene {\em et.~al.}, {\it
  {On the IR behaviour of the Landau-gauge ghost propagator}},  {\em JHEP} {\bf
  0806} (2008) 099 [\href{http://arXiv.org/abs/0803.2161}{{\tt 0803.2161}}].

\bibitem{Pene:2009iq}
O.~Pene, P.~Boucaud, J.~P. Leroy, A.~Le~Yaouanc, J.~Micheli {\em et.~al.}, {\it
  {A Ghost Story: Ghosts and Gluons in the IR regime of QCD}},  {\em PoS} {\bf
  QCD-TNT09} (2009) 035 [\href{http://arXiv.org/abs/0911.0468}{{\tt
  0911.0468}}].

\bibitem{RodriguezQuintero:2010wy}
J.~Rodriguez-Quintero, {\it {On the massive gluon propagator, the PT-BFM scheme
  and the low-momentum behaviour of decoupling and scaling DSE solutions}},
  {\em JHEP} {\bf 1101} (2011) 105 [\href{http://arXiv.org/abs/1005.4598}{{\tt
  1005.4598}}].

\bibitem{Watson:2010cn}
P.~Watson and H.~Reinhardt, {\it {The Coulomb gauge ghost Dyson-Schwinger
  equation}},  {\em Phys. Rev.} {\bf D82} (2010) 125010
  [\href{http://arXiv.org/abs/1007.2583}{{\tt 1007.2583}}].

\bibitem{Mandula:1987rh}
J.~E. Mandula and M.~Ogilvie, {\it {The Gluon Is Massive: A Lattice Calculation
  of the Gluon Propagator in the Landau Gauge}},  {\em Phys. Lett.} {\bf B185}
  (1987) 127--132.

\bibitem{Bicudo:2010wi}
P.~Bicudo and O.~Oliveira, {\it {Gluon Mass in Landau Gauge QCD}},  {\em PoS}
  {\bf LATTICE2010} (2010) 269 [\href{http://arXiv.org/abs/1010.1975}{{\tt
  1010.1975}}].

\bibitem{Boucaud:2011eh}
P.~Boucaud, D.~Dudal, J.~P. Leroy, O.~Pene and J.~Rodriguez-Quintero, {\it {On
  the leading OPE corrections to the ghost-gluon vertex and the Taylor
  theorem}},  \href{http://arXiv.org/abs/1109.3803}{{\tt 1109.3803}}.

\bibitem{Papavassiliou:2011ty}
J.~Papavassiliou, {\it {Schwinger mechanism in QCD}},
  \href{http://arXiv.org/abs/1112.0174}{{\tt 1112.0174}}.

\bibitem{Aguilar:2011xe}
A.~C. Aguilar, D.~Ibanez, V.~Mathieu and J.~Papavassiliou, {\it {Massless
  bound-state excitations and the Schwinger mechanism in QCD}},
  \href{http://arXiv.org/abs/1110.2633}{{\tt 1110.2633}}.

\bibitem{Aguilar:2009nf}
A.~C. Aguilar, D.~Binosi, J.~Papavassiliou and J.~Rodriguez-Quintero, {\it
  {Non-perturbative comparison of QCD effective charges}},  {\em Phys. Rev.}
  {\bf D80} (2009) 085018 [\href{http://arXiv.org/abs/0906.2633}{{\tt
  0906.2633}}].

\bibitem{Cornwall:1981zr}
J.~M. Cornwall, {\it {Dynamical Mass Generation in Continuum QCD}},  {\em Phys.
  Rev.} {\bf D26} (1982) 1453.

\end{thebibliography}\endgroup


\end{document}